\documentclass{epl}

\title{Tetravalent Colloids by Nematic Wetting}
\shorttitle{Tetravalent Colloids by Nematic Wetting}
\author{M. Huber \inst{1,2} \and H. Stark \inst{1}\thanks{E-mail: 
\email{Holger.Stark@uni-konstanz.de}}}
\shortauthor{M. Huber \and H. Stark}
\institute{
  \inst{1} Universit\"at Konstanz, Fachbereich Physik, D-78457
           Konstanz, Germany \\
  \inst{2} Department of Physics and Astronomy, 
           University of British Columbia, 6224 Agricultural Road, 
           Vancouver, B.C. V6T1Z1, Canada 
}

\begin{document}

\maketitle

\begin{abstract}
In an elegant paper, D. Nelson suggested a method to produce
tetravalent colloids based on a tetrahedral configuration created on the
surface of a spherical particle. It
emerges from a two-dimensional nematic liquid crystal placed
on a sphere due to the presence of four 1/2 disclinations, i.e.,
topological defects in the orientational order\ \cite{Nelson2002}.
In this paper we show that such a tetrahedral configuration also
occurs in the wetting layers which form around spheres dispersed
in a liquid crystal above the nematic-isotropic phase transition.
Nematic wetting therefore offers an alternative route towards
tetravalent colloids.
\end{abstract}

Recently, D. Nelson proposed a method by which tetravalent colloids
could be produced\ \cite{Nelson2002}. These micron-sized particles would 
have four chemical linkers or DNA strands symmetrically attached to their 
surfaces. Similar to, e.g., carbon, silicon and germanium atoms with their
sp$^{3}$ hybridized chemical bonds, the tetravalent colloids could 
then arrange into, e.g., a colloidal crystal with a diamond lattice 
structure which is known to possess a large photonic band gap\ \cite{Ho1990}.

\begin{figure}
\onefigure[width=12.5cm]{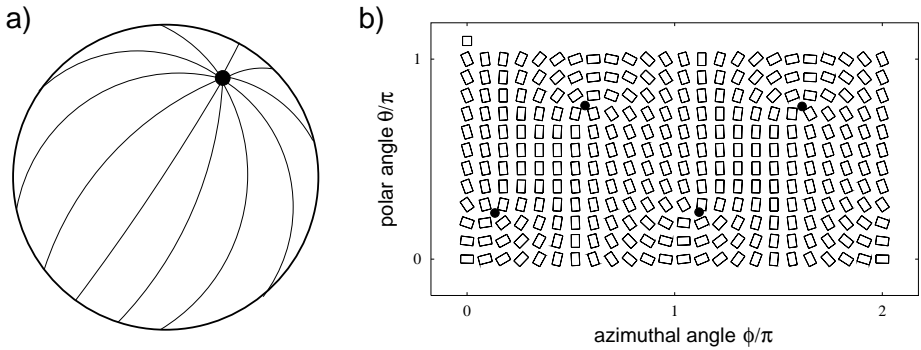}
\caption{Different configurations of a nematic liquid crystal on a
sphere. The pictures apply to both a pure two-dimensional nematic
and to biaxial nematic wetting layers. a) Boojum configuration with
two $+1$ disclinations at the north and south pole. The lines indicate 
the director field. b) Tetrahedral configuration visualized in the
angular space of spherical coordinates. The orientational order
in the sphere is indicated by rectangles. Four +1/2 disclinations
(see dots) are located on the vertices of a tetrahedron. As a reference,
the square in the upper left corner symbolizes a fully isotropic
order of the molecular axis.}
\label{f.1}
\end{figure}

To create the attachment sites for the chemical linkers, Nelson considered
the two-di\-men\-sio\-nal nematic liquid crystal phase on the particle surface
realized by a monolayer of elongated constituents such as gemini 
lipids\ \cite{Menger1993}, ABA triblock copolymers\cite{Matsen1990} or 
nanorods\ \cite{nanorods}.
One could first think that the rodlike particles on a sphere create
Mermin's boojums\ \cite{Mermin1977}, 
i.e., point defects with topological charge 1 situated
at the north and south pole of a sphere as illustrated in 
Fig.\ \ref{f.1}a). The lines indicate the director field. A vector
order parameter would indeed create such a configuration. The nematic
order, however, is described by an axis in space which allows the
boojums to split up into a pair of disclinations with charge 1/2. 
Lubensky and Prost showed that the ground state of a 2D nematic texture 
on a sphere consists of four 1/2 disclinations situated at the vertices 
of a tetrahedron\ \cite{Lubensky1992}. 
Fig.\ \ref{f.1}b) tries to illustrate such a 
configuration with tetrahedral symmetry in the angular space of 
spherical coordinates, i.e., polar angle $\theta$ versus azimuthal
angle $\phi$. For the moment, let the rectangles symbolize the 
nematic director. 
To visualize the boojum configuration in this representation, all the 
local directors would point along the vertical (lines of longitude).
Squares situated on the $\theta=0,\pi$ line (north and south pole) would 
indicate the isotropic order in the core of the boojums. To arrive at the 
tetrahedral configuration, the boojums split up into a total of four 
1/2 disclinations which move away from the north and south pole as 
indicated by the dots in Fig.\ \ref{f.1}b). If one is able to
attach chemical linkers selectively to the core of the disclinations,
then a tetravalent colloidal chemistry can be realized \cite{Nelson2002}.

Does the tetrahedral configuration survive when the colloid is suspended
in the nematic phase of a three-dimensional liquid crystal? The 
1/2 disclinations would extend as lines into the solvent. For 
topological reasons, they have to end either on neighboring particles or 
at the boundaries of the system. However, the resulting network of
defect lines carries a lot of free energy. It can be greatly reduced if
the particles realize the boojum configuration. The corresponding
+1 defect lines in the bulk nematic phase are not stable and 
``escape into the third dimension''\ \cite{escape}. 
This is indeed observed in nematic emulsions\ \cite{Poulin1998}.

We have recently looked at nematic wetting transitions and capillary
condensation in connection with colloids suspended in a liquid crystal
solvent above the nematic-isotropic phase transition\ \cite{wettspheres}.
So the interesting question arises if the tetrahedral configuration
occurs in nematic wetting layers which surround colloidal particles.
This would then suggest an alternative route towards tetravalent 
colloids. In answering the question, one first needs to establish
orientational order close to the colloid with a preferred axis 
parallel to its surface. Since the orientation of the preferred axis 
should be free to rotate within the surface we cannot just fix it by an
appropriate surface potential. All we can do is to favor planar ordering
of the liquid crystal molecules at the surface with an isotropic 
distribution of their axes in the surface. This uniaxial {\em oblate} order
is, however, energetically disfavored in the bulk since close to the 
phase transition the molecules already have a tendency to align parallel 
to each other, i.~e, to assume uniaxial {\em prolate} order. As a compromise, 
biaxial order close to the bounding surfaces occurs and introduces 
a preferred axis as required. The phase transition from uniaxial oblate to 
biaxial surface ordering is well studied in literature for planar
geometry\ \cite{Sluckin1985,Kothekar1994,planarwet}.
In the following, we readdress this problem to set the stage for 
the spherical geometry.
For the latter we then show that close to the nematic-isotropic phase 
transition, the biaxial ordering indeed leads to the tetrahedral 
configuration as in the two-dimensional
case. In addition, we find that the transition from oblate ordering
to the tetrahedral texture occurs via the boojum configuration 
which possesses a narrow stability region only.

To quantify the surface-induced orientational order, we use a traceless 
and symmetric second-rank tensor $Q_{ij}$, also called
alignment tensor. We perform our analysis with the help of
the Landau-Ginzburg-de\ Gennes free energy density in the bulk\ 
\cite{Gennes1971},
$f(Q_{ij}) = \frac{1}{2}a_{0}(T-T^{*}) Q_{ij}Q_{ij} - 
b Q_{ij} Q_{jk} Q_{ki} + c (Q_{ij}Q_{ij})^{2}
+ \frac{1}{2} L_{1} (Q_{ij,k})^{2}+\frac{1}{2} L_{2} (Q_{ij,j})^{2}$
where summation over repeated indices is implied and the symbol $,k$ 
means spatial derivative with respect to $x_{k}$. The first three terms 
describe the nematic-isotropic phase transition; $a_{0}$ and $c$ 
are positive constants and $T^{*}$ denotes the supercooling temperature 
of the isotropic phase. The fourth and fifth term penalize any non-uniform 
orientational order. Since we are interested in the basic features
of our system, we always choose $L_{1} = L_{2}$.
The anchoring of the molecules to a bounding surface 
is quantified by the Nobili-Durand free energy density,
$f_{S}(Q_{ij}) = \frac{W}{2} (Q_{ij} - Q_{ij}^{(0)})^{2}$,
where $W$ is the anchoring strength and $Q_{ij}^{(0)}$ the preferred
order parameter at the surface\ \cite{Nobili1992}. 
The number of parameters is reduced by using a rescaled order parameter 
$\mu_{ij} = Q_{ij}/s$ [$s=b/(\sqrt{6} c)$] and temperature 
$\tau=12 c a_{0}(T-T^{*}) / b^{2}$. Furthermore, all lengths
are given in terms of the particle radius $a$ ($\bar{\bm{r}} = \bm{r}/a$)
and the unit of the free energy is 
$\Delta f a^{3} = b^{4} a^{3} / (36 c^{3})$. We also introduce the
nematic coherence length at the nematic-isotropic phase transition,
$\xi_{r} = (12 c L_{1}/b^{2})^{1/2}$, and define the reduced 
anchoring strength as $\gamma = 6 c W/(b^{2} \xi_{r})$ so that the 
reduced free energy becomes
\begin{eqnarray}
F[\mu_{ij}(\bar{\bm{r}})] & = &
\int d^{3}\bar{r} \left( \frac{1}{4}\tau \mu_{ij} \mu_{ij} 
 - \sqrt{6} \mu_{ij} \mu_{jk} \mu_{kl} + (\mu_{ij} \mu_{ij})^{2}
 + \frac{1}{4} \frac{\xi_{r}^{2}}{a^{2}} 
[(\mu_{ij,k})^{2} + \textstyle{\frac{L_{2}}{L_{1}}}(\mu_{ij,j})^{2}]
\right) \nonumber\\
& & + \frac{\gamma}{2} \frac{\xi_{r}}{a} \int d^{2}\bar{r} 
\left(\mu_{ij} - \mu_{ij}^{(0)} \right)^{2} \enspace.
\label{1}
\end{eqnarray}
Note that in the planar case, we choose $a=\xi_{r}$ as the unit length.
Typical values for the nematic compound 5CB are 
$\Delta f = 8 \cdot 10^{5} \, \mathrm{erg/cm^{3}}$, 
$\xi_{r} = 10\,\mathrm{nm}$, and the temperature interval
$\Delta \tau = 1$ corresponds to $1.12 \, \mathrm{K}$.

A uniaxial order parameter $\mu_{ij} = \frac{3}{2} S (n_{i} n_{j} 
- \delta_{ij} / 3)$, where $\bm{n}$ is the nematic director and 
$\delta_{ij}$ the Kronecker symbol, minimizes the first three terms 
on the right-hand side of Eq.\ (\ref{1}). The bulk nematic-isotropic 
phase transition from $S=0$ to $S_{b} = \sqrt{6}/4$ occurs at 
$\tau_{c} = 1$, and $\tau^{\dagger} = 9/8$ is the superheating 
temperature of the nematic phase. The bulk exhibits prolate order 
since $S_{b} >0$, i.e., the rodlike molecules align along a common axis. 
In the following, we assume that bounding surfaces favor oblate order.
The preferred order parameter at the surface therefore assumes the 
form $\mu_{ij}^{(0)} = \frac{3}{2} S_{0} 
(\hat{\nu}_{i} \hat{\nu}_{j} - \delta_{ij} / 3)$ with $S_{0} < 0$.
In general, the order parameter can be biaxial. We visualize it by 
adding a term proportional to $\delta_{ij}$  and by calculating its
eigenvectors and eigenvalues whose directions and magnitudes then define
the edges of a cuboid. We find that in the planar or spherical geometry
one eigenvector always points along the surface normal or the radial
direction, respectively. So in Fig.\ \ref{f.1}b) we only see one 
face of the cuboid. To test for the biaxiality of $\mu_{ij}$, we 
introduce the measure\ \cite{Gramsbergen1986}
\begin{equation}
B=1-6\frac{(\mu_{ij}\mu_{jk}\mu_{ki})^{2}}{(\mu_{ij}\mu_{ij})^{3}} \enspace,
\end{equation}
which is zero for uniaxial tensors and assumes a maximal value of one.
An averaged degree of biaxiality is defined as
\begin{equation}
\bar{B} = \frac{\int d^{3}\bar{r} \mu_{ij}\mu_{ij} B}{
                \int d^{3}\bar{r} \mu_{ij}\mu_{ij}} \enspace.
\end{equation}
It weights the local $B$ with the degree of orientational order
given by $\mu_{ij}\mu_{ij}$ such that for $\mu_{ij} \rightarrow 0$ 
unphysically high local biaxialities do not contribute to $\bar{B}$.

Stable orientational textures correspond to a minimum of the 
free energy functional (\ref{1}). Its variation gives 
Euler-Lagrange equations for the five independent components of
$\mu_{ij}(\bar{\bm{r}})$. They are discretized and then solved numerically
by the standard Newton-Gauss-Seidel relaxation method using appropriate
starting configurations\ \cite{Press1992}. The planar geometry is just a 
one-dimensional problem with the bounding surface at $\bar{z}=0$.
To deal with the infinite half space, we introduced a new coordinate
$\rho=\exp(-\bar{z})$. Since close to the bulk phase transition
the biaxial wetting layer exhibits complete wetting\ \cite{Kothekar1994},
we also used an adaptive grid to resolve the interface between isotropic and 
orientational order. Oblate and biaxial wetting layers were distinguished 
by the biaxiality parameter $B$ at the surface. The wetting layers close to
a sphere were determined in a spherical coordinate system using the
local coordinate basis to define the alignment tensor 
$\mu_{ij}(\bar{\bm{r}})$. We replaced the radial coordinate $\bar{r}$
by $\rho=\exp[-(\bar{r}-1)a/\xi_{r}]$ to account for the infinite space
around the sphere. The Euler-Lagrange equations were formulated with the
help of differential geometry. Phase transitions between
different wetting layers were monitored by the averaged biaxiality
$\bar{B}$ and by the behavior of the free energy which we calculated
numerically by applying  Simpson's integration rule to Eq.\ (\ref{1}).

We add some comments about the possible orientational textures close
to a sphere. It has already been noted
that one eigenvector of the tensor
order parameter always points along the radial direction. From the 
remaining eigenvectors, we take the one with the largest eigenvalue 
as a director. Close to the particle it can then form the {\em splay} 
dominated boojum of Fig.\ \ref{f.1}a) or the {\em bend} dominated 
version where the director always points along the lines of latitude
of the sphere. Within our numerical accuracy, we observe that both 
configurations have the same free energy. This is due to the fact 
that in the Landau-De Gennes
theory splay and bend deformations are energetically degenerate.
The same applies to the bend dominated tetrahedral configuration
of Fig.\ \ref{f.1}b) and its splay dominated counterpart which follows
from Fig.\ \ref{f.1}b) by rotating all rectangles by $90^{\circ}$.
Note that Fig.\ \ref{f.1}b) illustrates the orientational order a 
distance of $0.05a$ away from the particle surface. We found that
around this distance the biaxiality of the orientational order is the 
largest. Topological defects in biaxial nematics are more complex than
in the uniaxial case\ \cite{biaxial}. However, since one eigenvector 
always points along the radial direction, we do not have to worry 
about these complications and can basically work with the director 
introduced above to identify the defects.

\begin{figure}
\onefigure[width=6.5cm]{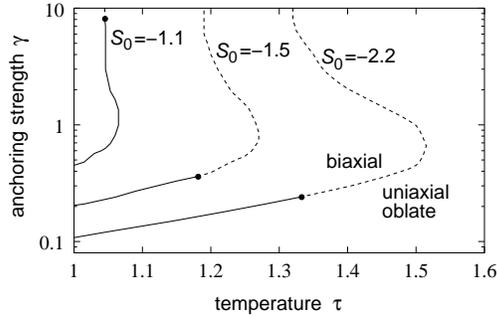}
\caption{Phase diagram for the planar geometry in terms of temperature 
$\tau$ and reduced anchoring strength $\gamma$; the parameter of the curves
is the preferred surface order parameter $S_{0}$. First and second-order phase 
transitions between uniaxial oblate and biaxial wetting layers are 
indicated by solid and dashed lines, respectively. The dot means 
tricritical point.}
\label{f.2}
\end{figure}

Before continuing the discussion of the spherical case, we
summarize our results for the planar geometry in Fig.\ \ref{f.2}.
We show a phase diagram in terms of temperature $\tau$ versus reduced
anchoring strength $\gamma$, the preferred surface order parameter $S_{0}$
is fixed for each transition line. 
Above $S_{0} \approx -1$ only the uniaxial oblate 
phase exists. For smaller $S_{0}$, a biaxial phase appears at 
temperatures close to the bulk phase transition ($\tau=1$) and for 
sufficiently large anchoring strength. Decreasing $S_{0}$ widens the
existence region of the  biaxial phase. The transition from the 
uniaxial oblate to the biaxial wetting layer is either of first 
(solid line) or second (dashed line) order. The transition lines 
meet in a tricritical point (dot). For large $\gamma$, the lines 
approach a constant temperature $\tau > 1$ (note the logarithmic scale for 
$\gamma$). Our results on a single transition line agree with earlier 
findings where a one-parameter surface potential with just a surface 
ordering field was used\ \cite{Kothekar1994}. With our more realistic 
surface potential, we show in addition that the biaxial wetting layer 
only appears for sufficently small $S_{0}$.

\begin{figure}
\onefigure[width=13.5cm]{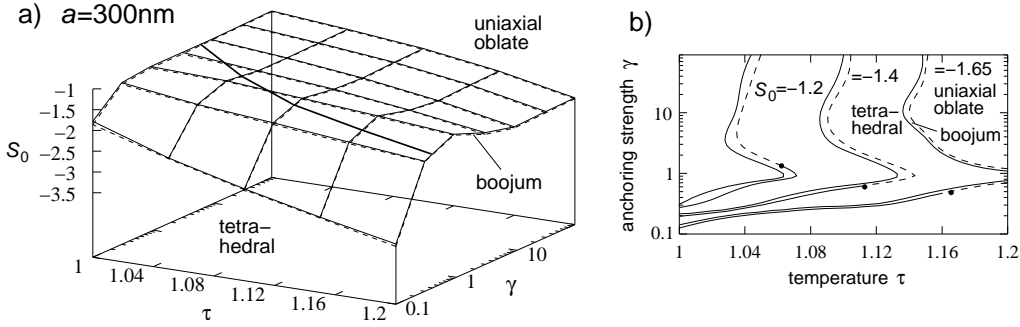}
\caption{a) Three-dimensional phase diagram of the different wetting layers
for $300\,\mathrm{nm}$ particles. Surface with solid grid lines: transition 
uniaxial oblate-boojum, surface with dashed grid lines (hardly visible): 
first-order transition boojum-tetrahedral. The solid line on the upper surface is 
a tricritical line. It separates a first-order (smaller $\gamma$) from 
a second-order (larger $\gamma$) transition. Note the very narrow 
existence region of the boojum. b) Contour plots of the three-dimensional 
phase diagram for various $S_{0}$.}
\label{f.3}
\end{figure}

Based on the knowledge of the planar case, we now address the spherical
geometry. For $300\, \mathrm{nm}$ particles ($a/\xi_{r}=30$), 
Fig.\ \ref{f.3}a) presents 
a three-dimensional phase diagram in terms of temperature $\tau$, 
reduced anchoring strength $\gamma$, and preferred order paramter $S_{0}$
locating the different types of wetting layers. The phase diagram
is the result of extensive numerical calculations which determined
the three-dimensional order parameter field close to the particle.
They were performed on a PC cluster and would take a few years
%
%
on a typical state-of-the art PC. As in the planar case, the uniaxial 
oblate wetting layer is always stable for $0 < S_{0} \le -1$. 
Decreasing $S_{0}$ induces first a phase transition to a boojum 
configuration for sufficiently large $\gamma$ and then to the tetrahedral 
wetting layer. Note that the stability region of the boojum is restricted 
to a very narrow $S_{0}$ intervall and, therefore, hardly visible in 
Fig.\ \ref{f.3}a). The solid line located on the upper surface indicates 
a line of tricritical points. 
To the left of this line, the transition from oblate ordering to the 
boojum is of first order and to the right it is of second order.
In contrast, the transition from the boojum to the tetrahedral
configuration is always discontinuous. We never found that the 
boojum evolves continuously into the tetrahedral configuration
via an asymmetric arrangement of the 1/2 disclinations. One could
speculate about the narrow stability region of the boojum configuration.
After all, as noted above, in the bulk nematic phase the corresponding
$+1$ disclination lines are not stable and ``escape into the third 
dimension'' to reduce the free energy. This, however, cannot occur in 
our case since for the biaxial order parameter the equivalent
$+1$ disclination line is actually stable\ \cite{biaxial}. So the
system reduces the free energy by adopting the tetrahedral configuration.

\begin{figure}
\onefigure[width=13.5cm]{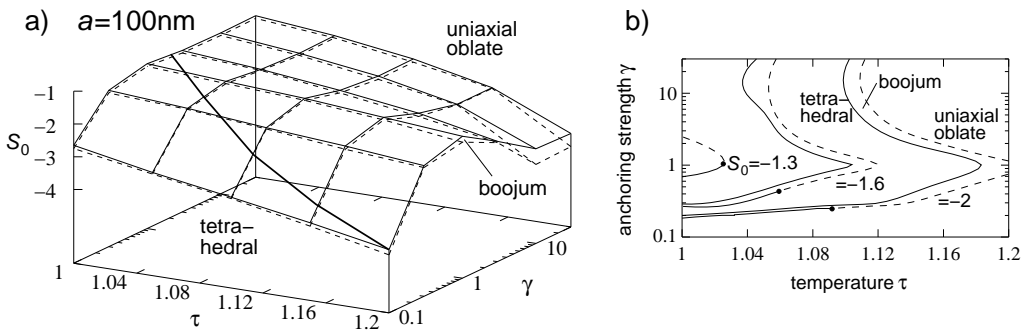}
\caption{The same phase diagrams as in Fig.\ \ref{f.3} but now for
$100\,\mathrm{nm}$ particles.}
\label{f.4}
\end{figure}

In Fig.\ \ref{f.3}b) we draw contour lines of the surfaces of the
three-dimensional phase diagram for different values of 
$S_{0}$ using the spline function of the graphical program gnuplot.
The resulting two-dimensional phase diagrams resemble very much 
the one in the planar case (see Fig.\ \ref{f.2}) with the exception 
that for decreasing temperature, we have a sequence of three phases as 
indicated for $S_{0}=-2$. 
As before, the solid and dashed lines mean first and second-order 
transitions, respectively, and the dots show the tricritical points.
To account for size effects, we also studied particles with radius
$a=100\,\mathrm{nm}$ ($a/\xi_{r}=10$). The phase diagrams in 
Fig.\ \ref{f.4} qualtitatively look the same as for the larger
particles. Comparing the two-dimensional diagrams of the planar case
and the two particle sizes for a fixed $S_{0}$ , one clearly recognizes 
that with increasing curvature the temperature range where the biaxial
wetting layers exist becomes smaller. So curvature suppresses the
biaxial phases. Surprisingly, the rough location of the transition 
lines with respect to the anchoring strength $\gamma$ does not seem 
to be affected by curvature.

How reasonable are the rescaled parameters of the surface potential 
where the biaxial wetting layers are observed? Using
Landau parameters from 5CB ($b=0.7\cdot10^{7}\,\mathrm{erg/cm^{3}}$,
$c=0.5\cdot10^{7}\,\mathrm{erg/cm^{3}}$) \cite{Coles1978} and 
$\xi_{r}=10\,\mathrm{nm}$, we arrive at $\gamma = W / 
\mathrm{1\frac{erg}{cm^{3}}}$. So the well achievable anchoring strength
$W = 1\mathrm{erg/cm^{3}}$ gives the right order of magnitude for 
$\gamma$. According to the Landau-de Gennes theory, the rescaled bulk nematic
order parameter at the phase transition is $S_{b} = 0.61$ which corresponds
to the measured typical value of $S_{mb}=0.3$ for the Maier-Saupe order 
parameter $S_{m} = \langle 3 \cos\theta - 1\rangle /2$\ \cite{Gennes1993a}. 
For a perfectly aligned nematic (prolate order) $S_{m}=1$; the perfect 
oblate order gives $S_{m}=-1/2$. This means that in our scaling
the minimum value of
$S_{0}$ in the surface potential is $S_{0} = -S_{b}/S_{mb}/2 = -1.02$. 
This value is further decreased for smaller $S_{mb}$, so that our predicted
transitions from the oblate to the biaxial surface phases, especially
the tetrahedral configuration, should be observable.
%
%
%

Finally, we add some comments about the influence of fluctuations on our
phase diagrams. It was argued that in the planar geometry, the 
transition from the biaxial to the oblate wetting layer is governed by
defect unbinding, i.e., by the Berezinskii-Kosterlitz-Thouless (BKT)
mechanism\ \cite{Sluckin1985,Kothekar1994,planarwet}.
As a result, the second-order transition line is shifted to smaller 
temperatures (see e.g.\ Ref.\ \cite{Kothekar1994}). The treatment of
fluctuations is beyond the scope of this paper. However, it seems possible
that under the BKT mechanism the boojum configuration vanishes completely
from the phase diagrams in Fig.\ \ref{f.3} and \ref{f.4} due to its
very narrow existence region. In an elaborate and elegant treatment,
Nelson finds that the tetrahedral configuration of the two-dimensional 
nematic is stable against thermal fluctuations deep in the nematic phase, 
whereas close to the phase transition the fluctuations are fairly 
large\ \cite{Nelson2002}. How these results apply to our studies
is not completely clear. It certainly depends on how strong the
biaxial ordering is developed in the tetrahedral configuration.

In conclusion, we have shown that nematic wetting layers around spherical
particles possess indeed a tetrahedral configuration where 1/2 
disclinations are located at the vertices of a tetrahedron. This offers
an interesting route towards tetravalent colloids. Planar anchoring, 
necessary for the realization of this tetrahedral configuration, can be
realized in inverted nematic emulsions\ \cite{Poulin1998}, or in lyotropic
liquid crystals conmposed of rodlike surfactant micelles\ \cite{Poulin1999}.
A possible method to detect nonuniform wetting layers is depolarized 
light scattering\ \cite{Berne1976}. Furthermore, the Brownian rotational 
motion of the particles should be measurable by dynamic light scattering\ 
\cite{Mertelj2002}.

\acknowledgments
The authors thank W. Poon for helpful discussions and the Deutsche
Forschungsgemeinschaft for financial support under Grant No. Sta 352/5-2
and within the International Graduate College ``Soft Matter''.






\begin{thebibliography}{10}



\bibitem{Nelson2002}
\Name{Nelson D.~R.}
\REVIEW{Nano Letters}{2}{2002}{1125}.

\bibitem{Ho1990}
\Name{Ho K.~M., Chang C.~T. \and Soukoulis C.~M.}
\REVIEW{Phys.\ Rev.\ Lett.}{65}{1990}{3152}.

\bibitem{Menger1993}
\Name{Menger F.~M. \and Littau C.~A.}
\REVIEW{J.\ Am.\ Chem.\ Soc.}{115}{1993}{10083}.

\bibitem{Matsen1990}
\Name{See, e.g., Matsen M.~W. \and Schick M.}
\REVIEW{Macromolecules}{27}{1990}{187}.

\bibitem{nanorods}
\Name{Nikoobakht B., Wang Z.~L. \and El-Sayed M.~A.}
\REVIEW{J.\ Phys.\ Chem.\ B}{104}{2000}{8635};
\Name{Kim F., Kwan S., Akana J. \and Yang P.}
\REVIEW{J.\ Am.\ Chem. Soc.}{123}{2001}{4360};
\Name{Li L., Walda J, Manna L. \and Alivisatos A.P.}
\REVIEW{Nano Letters (Communication)}{2}{2002}{557}.

\bibitem{Mermin1977}
\Name{Mermin N.} in
\Book{Quantum Fluids and Solids}
\Editor{S.~B. {Trickey}, E.~D. {Adams} \and J.~W. {Dufty}}
\Publ{Plenum Press, New York}
\Year{1977}
\Pages{3}{22}.

\bibitem{Lubensky1992}
\Name{Lubensky T.C. \and Prost J.}
\REVIEW{J. Phys. II (France)}{2}{1992}{371}.

\bibitem{escape}
\Name{Cladis P.~E. \and Kl{\'e}man M.}
\REVIEW{J.\ Phys.\ (Paris)}{33}{1972}{591};
\Name{Williams C., Piera\'nski P., and Cladis P.~E.}
\REVIEW{Phys.\ Rev.\ Lett.}{29}{1972}{90};
\Name{Meyer R.~B.}
\REVIEW{Philos.\ Mag.}{27}{1973}{405}.

\bibitem{Poulin1998}
\Name{Poulin P. \and Weitz D.~A.}
\REVIEW{Phys.\ Rev.~E}{57}{1998}{626}.

\bibitem{wettspheres}
\Name{Fukuda J., Stark H. \and Yokoyama H.}
\REVIEW{Phys.\ Rev.\ E}{69}{2004}{021714};
\Name{Stark H., Fukuda J. \and Yokoyama H.}
\REVIEW{Phys.\ Rev.\ Lett.}{92}{2004}{205502}.

\bibitem{Sluckin1985}
\Name{Sluckin T.~J. \and Poniewierski A.}
\REVIEW{Phys.\ Rev.\ Lett.}{55}{1985}{2907}.

\bibitem{Kothekar1994}
\Name{Kothekar N., Allender D.~W. \and Hornreich R.~M.}
\REVIEW{Phys.\ Rev.\ E}{49}{1994}{2150}.

\bibitem{planarwet}
\Name{Hornreich R.~M, Kats E.~I. \and Lebedev V.~V.}
\REVIEW{Phys.\ Rev.\ A}{46}{1992}{4935};
\Name{Y. L'vov, Hornreich R.~M. \and Allender D.~W.}
\REVIEW{Phys.\ Rev.\ E}{48}{1993}{1115};
\Name{Seidin R., Hornreich R.~M. \and Allender D.~W.}
\REVIEW{Phys.\ Rev.\ E}{55}{1997}{4302}.

\bibitem{Gennes1971}
\Name{de Gennes P.~G.}
\REVIEW{Mol.\ Cryst.\ Liq.\ Cryst.}{12}{1971}{193}.

\bibitem{Nobili1992}
\Name{Nobili M. \and Durand G.}
\REVIEW{Phys.\ Rev.~A}{46}{1992}{R6174}.

\bibitem{Gramsbergen1986}
\Name{Gramsbergen E.~F., Longa L. \and de Jeu W.~H.}
\REVIEW{Phys.\ Rep.}{135}{1986}{195}.

\bibitem{Press1992}
\Name{Press W.~H., Teukolsky S.~A., Vetterling W.~T.\and Flannery B.~P.}
\Book{{N}umerical {R}ecipes in {F}ortran: {T}he {A}rt of {S}cientific
  {C}omputing}
\Publ{Cambridge University Press, Cambridge}
\Year{1992}.

\bibitem{biaxial}
\Name{Toulouse G.}
\REVIEW{J.~Phys. (Paris)\ Lett.}{38}{1977}{L67};
\Name{Mermin N.~D.}
\REVIEW{Rev. Mod. Phys.}{51}{1979}{591};
\Name{Kleman M. \and Lavrentovich O.}
\Book{Soft Matter Physics}
\Publ{Springer, Berlin}
\Year{2001}

\bibitem{Coles1978}
\Name{Coles H.~J.}
\REVIEW{Mol.\ Cryst.\ Liq.\ Cryst.}{49}{1978}{67}.

\bibitem{Gennes1993a}
\Name{de Gennes P.~G. \and Prost J.}
\Book{The Physics of Liquid Crystals}
\Publ{Oxford Science Publications, Oxford}
\Year{1993}

\bibitem{Poulin1999}
\Name{Poulin P., Franc\`es N. \and Mondain-Monval O.}
\REVIEW{Phys.\ Rev.\ E}{59}{1999}{4384}

\bibitem{Berne1976}
\Name{Berne B.~J. \and Pecora R.}
\Book{Dynamic Light Scattering -- with
  applications to chemistry, biology, and physics}
\Publ{John Wiley, New York}
\Year{1976}

\bibitem{Mertelj2002}
\Name{Mertelj A., Arauz Lara J.~L., Maret G., Gisler T. \and Stark H.}
\REVIEW{Europhys. Lett.}{59}{2002}{337}.





\end{thebibliography}
\end{document}